%% file: ms.tex
\documentclass{osa-article}

\journal{osajournal}

\articletype{Research Article}

\begin{document}

\title{Controlling spatial coherence with an optical complex medium}

\author{Alfonso Nardi,\authormark{1,*} Felix Tebbenjohanns,\authormark{1,2} Massimiliano Rossi,\authormark{1} Shawn Divitt,\authormark{3} Andreas Norrman,\authormark{1,4} Sylvain Gigan,\authormark{5} Martin Frimmer,\authormark{1} Lukas Novotny\authormark{1,6}}

\address{\authormark{1}Photonics Laboratory, ETH Zürich, Zürich, Switzerland\\
\authormark{2}Currently with the Department of Physics, Humboldt-Universität zu Berlin, 10099 Berlin, Germany\\ \authormark{3}Gaithersburg, MD USA\\ \authormark{4}Institute of Photonics, University of Eastern Finland, P.O. Box 111, FI-80101 Joensuu, Finland\\ \authormark{5}Laboratoire Kastler Brossel, École Normale Supérieure-Université PSL, CNRS, Sorbonne Université, College de France, Paris 75005, France\\ \authormark{6}Quantum Center, ETH Zürich, Zürich, Switzerland}

\email{\authormark{*}anardi@ethz.ch} 

\begin{abstract}
Control over optical spatial coherence is a key requirement in diverse applications including imaging, optical trapping, and communications.
Current approaches to controlling spatial coherence are constrained by speed or limited to a single pair of optical fields.
Here, we propose a method to achieve single-shot control of the spatial coherence between an arbitrary number of fields.
Our method employs a multi-port linear optical device, which we realize by shaping the wavefront of the input light fields and transmitting them through a complex medium. 
To demonstrate the capabilities of our method, we experimentally realize a $3 \times 3-$port system and use it to generate three output beams with desired mutual correlations.
\end{abstract}

\section{Introduction}
Coherence is a fundamental notion in the science of light and offers a central degree of freedom to manipulate electromagnetic radiation of diverse physical character \cite{mandel1995optical}.
The concept of spatial coherence, in particular, has had a profound impact on the foundations and development of modern optics \cite{wolf2007influence}, and it is manifested in countless technological applications, e.g., in imaging \cite{Ostrovsky2011, Redding2012}, tomography \cite{Baleine2004, Baleine2004b, Baleine2005}, beam propagation \cite{Gbur2002}, nanophotonics \cite{Norrman2016, CHEN2020105}, trapping \cite{Wang2007, Aun2012, Aun2013, Divitt2015}, and free-space optical communications \cite{Korotkova2004, Gbur2014}.
Nevertheless, the control of spatial coherence is not yet exploited to its full extent.
Researchers have developed several methods to control this coherence property, either via active devices such as spinning phase diffusers \cite{DeSantis1979}, spatial light modulators (SLMs) \cite{Hyde2015} and digital micromirror devices (DMDs) \cite{Rodenburg2013}, or via passive methods such as fine tuning of the optical path \cite{K.R.2019, Divitt2016, li2017strong}.
Yet, all these techniques have limitations on the attainable speed or are limited to a single pair of fields.

Recently, a new class of tools to control light has emerged that exploits the properties of optical complex media when combined with wavefront shaping devices.
A complex medium is an optical system that mixes the spatial and temporal degrees of freedom of an impinging field, resulting in a scrambled intensity distribution at its output \cite{goodman2020speckle}.
The extremely large number of internal degrees of freedom of a complex medium makes the output intensity pattern disordered, yet it remains deterministic.
Therefore, it is possible to fully characterize the effect of the propagation through the medium on an incident field with a linear transmission matrix (TM) \cite{Popoff2010}.
Knowledge of the TM allows complex media to be used to perform a variety of tasks, once combined with programmable modulators.
Applications include the control of different properties of light, e.g., intensity \cite{Vellekoop2010, VanPutten2011, Mounaix2017}, polarization \cite{Xiong2018, Guan2012} and spectrum \cite{Andreoli2015, Mounaix2016, Mounaix2018, Boniface2021}.
In particular, complex media have been proposed as a compact, highly-dimensional multi-port device \cite{Matthes:19}, e.g., to perform quantum operations \cite{Wolterink2016, Leedumrongwatthanakun2020}.
The use of these media in combination with wavefront shaping presents a potential alternative to other platforms, in free space \cite{Reck1994} or integrated optics \cite{Miller2013, Carolan2015, Ribeiro2016}, which suffer from scalability issues.
Interestingly, even though both random diffusers and wavefront shaping devices, such as SLMs and DMDs, have been used for the control of the spatial coherence \cite{DeSantis1979, Hyde2015, Rodenburg2013}, they have not been employed in combination to overcome the previous limitations.

In this work, we propose a technique to control the correlations between an arbitrary number of field pairs, based on a linear transformation applied to $n$ mutually incoherent input fields.
Employing optical coherence theory, we first derive a general expression for the linear transformation that yields access to such coherence control.
The linear transformation is then experimentally implemented with a complex medium in combination with SLM-based wavefront shaping.
As a proof of principle, we realize a $3 \times 3-$port device and show that it generates any combination of mutual correlations, within the technical limitations.

\section{Theory}\label{sec:theory}

In this section, after providing some background concepts, we show how to generate a set of $n$ fields with precisely controlled mutual correlations by applying an $n \times n$ linear transformation to a set of $n$ mutually incoherent input fields.

\subsection{Basic definitions}

Let the complex analytic signals $E_1,\ldots,E_n$ represent random, statistically stationary, quasimonochromatic electric fields at $n$ different points in space. The spatial correlation between two fields $E_i$ and $E_j$, with $i,j\in\{1,\ldots,n\}$, can be characterized via the mutual (or complex) degree of coherence \cite{BornWolf}
\begin{equation}
\gamma_{ij}=\frac{\langle E_{i}E_{j}^{\ast}\rangle}{\sqrt{\langle|E_{i}|^{2}\rangle\langle|E_{j}|^{2}\rangle}}\;,
\end{equation}
where the angle brackets stand for ensemble or time average (equivalent with ergodic and stationary fields). For each field-field pair we have that $0\leq|\gamma_{ij}|\leq1$, with the upper and lower limits corresponding to full coherence and full incoherence, respectively, while the intermediate values represents partial coherence. By introducing the column vector $\boldsymbol{E} = \left[ E_1/\sqrt{I_1}, E_2/\sqrt{I_2},\ldots, E_n/\sqrt{I_n} \right]^\intercal$, with $I_i = \langle |E_i|^2 \rangle$, we can collect all the degrees of coherence into an $n\times n$ spatial coherence matrix (abbreviated to \textit{coherence matrix} from now on)
\begin{equation}
     \mathbb{K} = \langle \boldsymbol{E} \boldsymbol{E}^\dagger \rangle\;,
\end{equation}
where the dagger denotes conjugate transpose. The matrix $\mathbb{K}$ thus contains all the information about the spatial coherence in the system. The diagonal elements are given by the self degrees of coherence $\gamma_{ii}$, which are always equal to $1$, while the off-diagonal terms are the mutual degrees of coherence $\gamma_{ij}$.
The coherence matrix is also known as a statistical correlation matrix, which is a normalized covariance matrix, and must be Hermitian and positive semi-definite \cite{Devroye1986}.

To characterize the overall spatial coherence of the system, we employ the measure \cite{gil2020sources}
\begin{equation}\label{S}
    \mathcal{S} = \frac{n}{n-1}\left[ \frac{ \mathrm{tr}\left(\mathbb{K}^2 \right)}{\left(\mathrm{tr}\mathbb{K}\right)^{2}} - \frac{1}{n} \right]\;,
\end{equation}
where tr stands for matrix trace. The quantity $\mathcal{S}$ is the distance between the coherence matrix $\mathbb{K}$ and the $n\times n$ identity matrix $\mathbb{I}$, with the scaling in Eq.~(\ref{S}) chosen such that $0\leq\mathcal{S}\leq1$. The upper bound $\mathcal{S}=1$ is saturated exclusively when all fields are mutually fully coherent ($|\gamma_{ij}|=1$), hence corresponding to a spatially completely coherent system. The lower bound $\mathcal{S}=0$, on the other hand, is met only when all fields are mutually fully incoherent ($|\gamma_{ij}|=0$, for $i \neq j$), in which case the whole system is spatially completely incoherent and the coherence matrix is equal to the identity matrix ($\mathbb{K}=\mathbb{I}$).

\subsection{Coherence control with linear transformation} \label{sec:theory_coherence_control}

Let us now consider a vector $\boldsymbol{E}_{\mathrm{in}}$ that represents $n$ mutually incoherent input fields. Since in this case all mutual degrees of coherence are zero, whereupon the overall spatial coherence of the whole system is also zero, the input coherence matrix obeys $\mathbb{K}_{\mathrm{in}} = \langle \boldsymbol{E}_{\mathrm{in}} \boldsymbol{E}_{\mathrm{in}}^\dagger \rangle = \mathbb{I}$. We combine the input fields via a linear transformation $\hat{T}$, according to
\begin{equation}
    \boldsymbol{E}_{\mathrm{out}} = \hat{T}\boldsymbol{E}_{\mathrm{in}}\;,
\end{equation}
where $\boldsymbol{E}_{\mathrm{out}}$ is the vector describing the output fields. The output coherence matrix is then given by
\begin{equation}
    \mathbb{K}_{\mathrm{out}} = \langle \boldsymbol{E}_{\mathrm{out}}\boldsymbol{E}_{\mathrm{out}}^\dagger \rangle
    = \langle \hat{T}\boldsymbol{E}_{\mathrm{in}}\boldsymbol{E}_{\mathrm{in}}^\dagger\hat{T}^\dagger \rangle  \;.
\end{equation}
Using the fact that $\hat{T}$ is deterministic and the inputs are mutually incoherent, we obtain
\begin{equation}\label{eq:KtoT}
    \mathbb{K}_{\mathrm{out}} = \hat{T} \langle \boldsymbol{E}_{\mathrm{in}}\boldsymbol{E}_{\mathrm{in}}^\dagger \rangle \hat{T}^\dagger =
    \hat{T}  \mathbb{K}_{\mathrm{in}} \hat{T}^\dagger = 
    \hat{T}\hat{T}^\dagger\;.
\end{equation}
Therefore, it is possible to generate an arbitrary output coherence matrix upon choosing a linear transformation which fulfills
\begin{equation}\label{eq:Texpression}
    \hat{T} = \sqrt{\mathbb{K}_{\mathrm{out}}}\;,
\end{equation}
where the square root is the principal square root of the matrix. The positive semi-definiteness of the coherence matrix ensures the existence of such a linear transformation~\cite{Devroye1986}.
We note that the assumption of mutually incoherent input fields is not necessary to control the output coherence, yet it simplifies the treatment, as indicated by Eqs.~(\ref{eq:KtoT}) and (\ref{eq:Texpression}).

We observe from Eq.~\eqref{eq:KtoT} that under unitary transformations, $\hat{T}\hat{T}^\dagger = \hat{T}^\dagger\hat{T} = \mathbb{I}$, the output coherence matrix always obeys $\mathbb{K}_{\mathrm{out}}=\mathbb{K}_{\mathrm{in}}=\mathbb{I}$ for an incoherent input system.
The control of the output coherence thus relies on the nonunitary character of the chosen transformation. In practice, we implement the desired transformation using a multi-port linear optical device, as we describe in the following section.

\section{Implementation}\label{sec:implemetation}
We implement a multi-port linear optical device by transmitting wavefront-shaped light through a complex medium.
To use the medium to perform the desired transformation, we need to characterize its transmission matrix. 
To do that, we use a phase-only SLM to inject light into the medium in a well defined input basis, and we measure the output speckle pattern for each vector of the basis \cite{Popoff2010}.
The number of degrees of freedom of the scattering layer (given by the number of the propagating modes supported by the material) is practically unlimited, hence the dimensionality of the TM is only limited by the number of pixels of the SLM and of the camera \cite{Rotter2017}, both of which are typically on the order of one million.
From the knowledge of this large fixed random matrix, we can employ the SLM in combination with the complex medium to implement a smaller but reconfigurable linear transformation \cite{Matthes:19}.
Importantly, since the whole scattering matrix is too large to be fully characterized with the limited number of pixels of the SLM, all the uncontrolled modes can be considered lossy channels.
These channels enable the implementation of non-unitary transformations with our compact system, comprising the SLM and the complex medium, hence allowing the control of the coherence matrix of the output fields.
We note that, in integrated or standard free-space optics, realizing non-unitary transformations requires a large number of optical components, many of which are employed only to implement lossy channels \cite{Tischler2018}.

\subsection{Multi-port linear device}\label{subsec:linear_port}
Here, we describe how to obtain a programmable linear transformation for coherence control using the system of SLM and complex medium.
For illustration purposes, we focus on the simple case of a linear $3\times3-$port device, which we implement experimentally.

Let us consider three mutually incoherent light beams $E^{\mathrm{in}}_1$, $E^{\mathrm{in}}_2$ and $E^{\mathrm{in}}_3$.
The three non-overlapping beams reach different regions of a phase-only SLM.
For each region, $N$ segments of the SLM modulate the field locally, effectively generating $N$ spatially separated modes with controlled phase.
Thus, each input $E^{\mathrm{in}}_i$ undergoes a transformation $\hat{T}_{\mathrm{SLM}, i}$ with dimensions $N \times 1$.
Next, the three sets of $N$ modes enter the complex medium.
The output intensity emanating from a complex medium typically forms a disordered interference pattern, known as speckle pattern, resulting from the mixing of the input modes.
In the case under analysis, the three sets of modes entering the medium are mutually incoherent, so they do not interfere, thus leading to the sum of three independent speckle patterns.
Hence, we characterize the effect of the medium with three distinct TMs, which we denote as $\hat{T}_{\mathrm{CM}, i}$.
The dimension of each $\hat{T}_{\mathrm{CM}, i}$ is $M \times N$, since they connect each set of $N$ input modes to $M$ output speckles.
We want to control only three modes out of the many ones at the output. 
These three modes should then enclose the largest amount of the output power.
Therefore, we apply a projection $\hat{P}$ (of size $3 \times M$) to each $\hat{T}_{\mathrm{CM}, i}$ to select only the output speckles we are interested in, while zeroing out the intensity of the rest.

The overall operation transforms each input beam into three outputs, according to the relation $[ E^{\mathrm{out}}_{1,i},  E^{\mathrm{out}}_{2,i}, E^{\mathrm{out}}_{3,i} ]^\intercal = \hat{P}\hat{T}_{\mathrm{CM}, i} \hat{T}_{\mathrm{SLM}, i} E^{\mathrm{in}}_i$, for $i = 1,2,3$.
We then sum the independent speckle patterns to get the final result:
\begin{equation}
    \begin{bmatrix}
        E^{\mathrm{out}}_1 \\ E^{\mathrm{out}}_2 \\ E^{\mathrm{out}}_3
    \end{bmatrix}
    = \sum_{i=1}^3 \hat{P}\hat{T}_{\mathrm{CM}, i} \hat{T}_{\mathrm{SLM}, i} E^{\mathrm{in}}_i    = 
    \begin{bmatrix}
        t_{11} & t_{12} & t_{13} \\
        t_{21} & t_{22} & t_{23} \\
        t_{31} & t_{32} & t_{33} 
    \end{bmatrix}
    \begin{bmatrix}
        E^{\mathrm{in}}_1 \\ E^{\mathrm{in}}_2 \\ E^{\mathrm{in}}_3
    \end{bmatrix}\;,
\end{equation}
where
\begin{equation} \label{eq:coeff_implementation}
    \hat{P}\hat{T}_{\mathrm{CM}, i} \hat{T}_{\mathrm{SLM}, i} = 
    \begin{bmatrix}
        t_{1i} \\ t_{2i} \\ t_{3i}
    \end{bmatrix}\;.
\end{equation}
Since we know from Eq.~\eqref{eq:Texpression} the target coefficients $t_{ij}$ to obtain the desired output coherence matrix, we only need to invert Eq.~\eqref{eq:coeff_implementation} to find the transformation to be implemented with the SLM.
In practice, we apply a phase conjugation, which has already been proven successful to focus light into few speckles \cite{Popoff2011b}.
Finally, we get the following relation:
\begin{equation}\label{eq:inversion}
    \hat{T}_{\mathrm{SLM},i} = \hat{T}_{\mathrm{CM}, i}^\dagger \hat{P}^\dagger \begin{bmatrix}
        t_{1i} \\ t_{2i} \\ t_{3i}
    \end{bmatrix}\;,
\end{equation}
which is the configuration that we encode into the SLM to implement the desired transformation.

\section{Experiment}\label{sec:Experiment}
\begin{figure}
    \centering
    \includegraphics[width=0.5\textwidth]{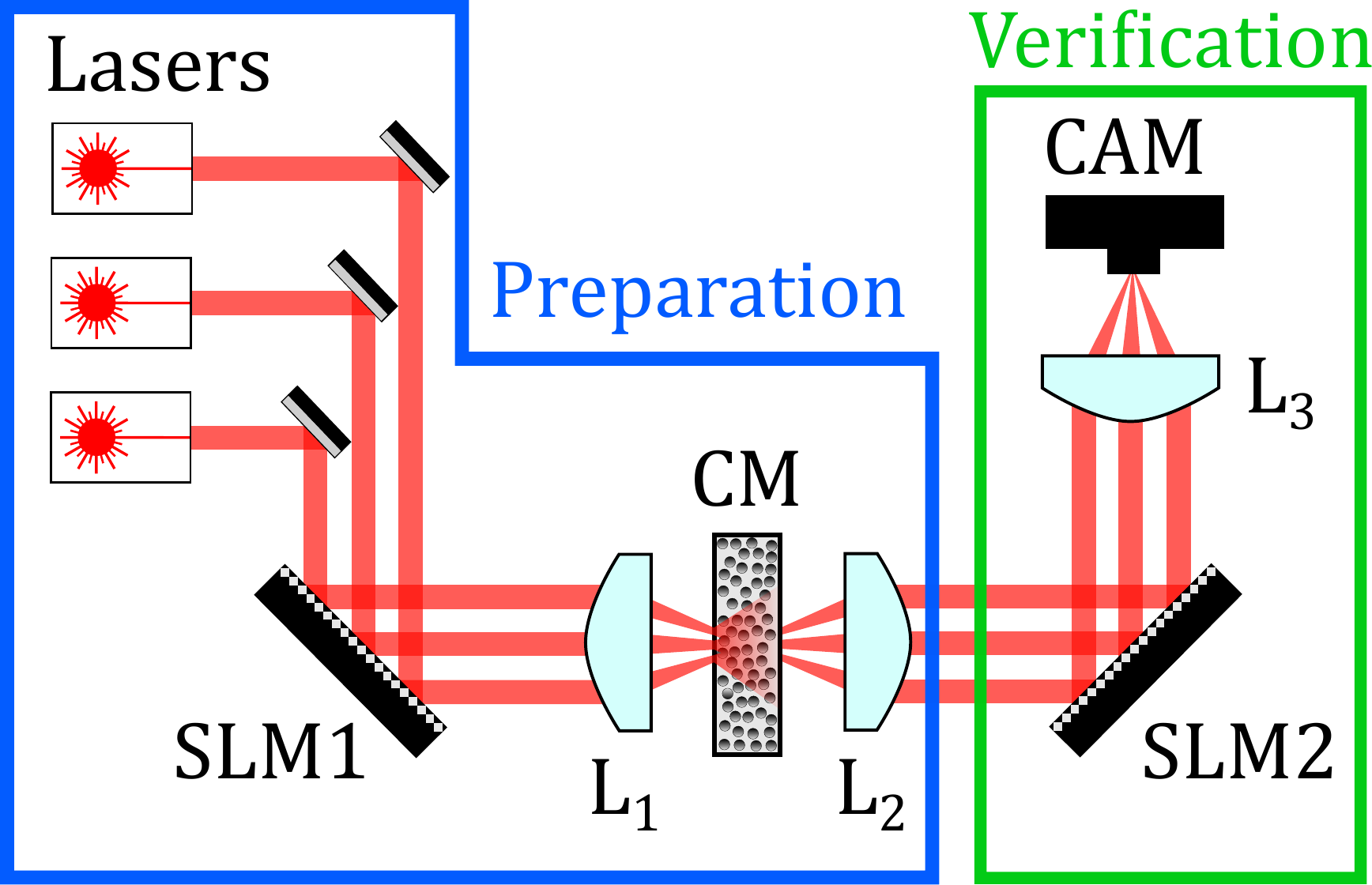}
    \caption{Experimental setup. 
    We employ three different lasers as mutually incoherent inputs. 
    The three lasers are modulated by a phase-only SLM, then they are focused onto a complex medium (ground glass diffuser) by a lens.
    The propagating beams are mixed by the complex medium and then collected by another lens.
    Through wavefront shaping, we obtain three output beams with the desired degree of coherence. 
    The three output beams are focused by a third lens and interfere in the camera plane.
    A second SLM is used to characterize the degree of coherence from the interference pattern. $L_1$, $L_2$, $L_3$: lenses; $\text{CM}$: complex medium; $\text{CAM}$: camera.}
    \label{fig:experimental_setup}
\end{figure}
In Fig.~\ref{fig:experimental_setup} we show the experimental setup, which comprises two main blocks.
The first one (preparation) generates three fields characterized by a programmed coherence matrix, while the second block (verification) verifies that the encoded degrees of coherence correspond to the desired ones. 

We use two $512\times512$ pixel spatial light modulators (Meadowlark Optics P512).
In the preparation stage, we use a first SLM (SLM1) to modulate three mutually incoherent input lasers (Thorlabs HRP050 and Meredith Instruments $633\,\mathrm{nm}$ HeNe lasers, and $\approx 650\,\mathrm{nm}$ FOSCO BOB-VFL650-10, see Supplement 1 for more details).
Next, we focus them onto an optical complex medium (ground glass diffuser, Thorlabs DG10-1500).
The SLM1 and the scattering medium together form the programmable multi-port linear device.
Through wavefront shaping, the light scattered by the medium and collected by a lens forms three output beams.

In the verification stage, we use a second SLM (SLM2) to modulate the phase of the beams before a lens. 
This allows us to control which beam is focused onto the camera plane and to which location.
From the interference patterns measured with the camera (Basler acA640-750um), we reconstruct the mutual degree of coherence.
In the following sections, we describe the procedures used to encode and measure the coherence matrix of the output fields.

\begin{figure*}
    \centering
    \includegraphics[width=\textwidth]{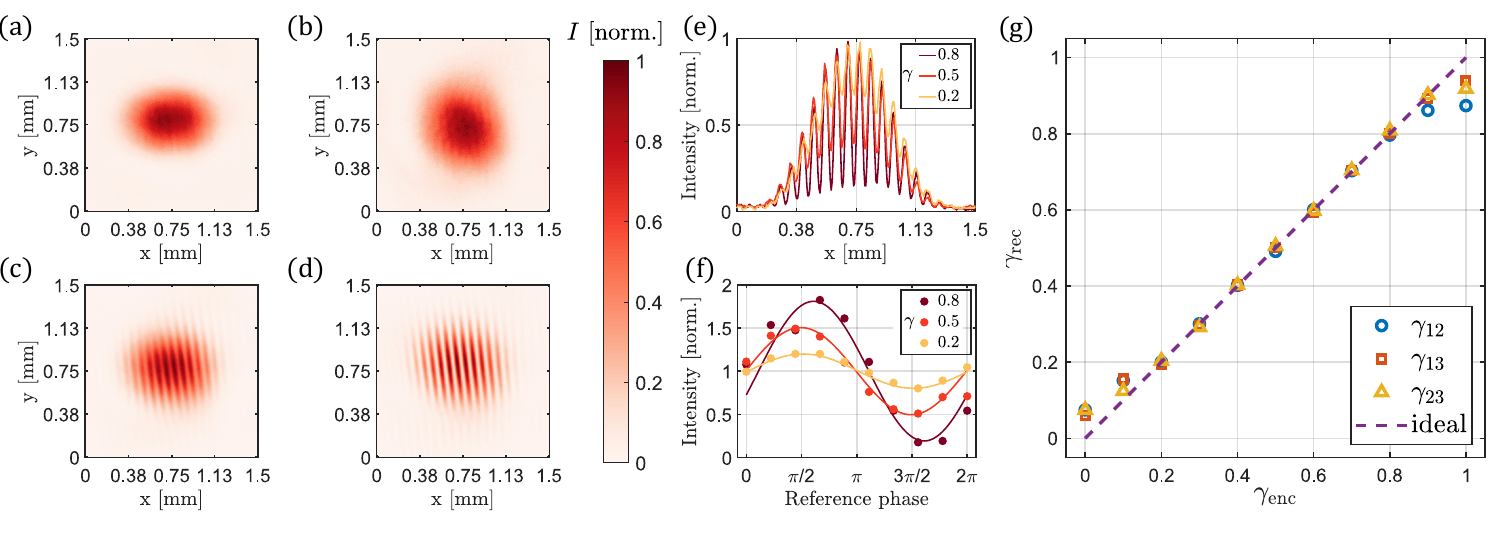}
    \caption{Reconstruction of the degree of coherence.
    (a, b) Intensity $I$ of the (a) first and the (b) second beam. 
    (c, d) Interference patterns for the degree of coherence ($\gamma$) equal to (c) 0.2 and (d) 0.8. 
    (e) Cross-sections of the intensity distributions for different values of $\gamma$.
    (f) Normalized intensity distribution at a fixed pixel as a function of the phase of one of the interfering beams (reference phase), which is swept from 0 to $2\pi$.
    The different coloured dots correspond to measurements taken for different degrees of coherence. 
    The solid lines are the cosine fits of the data points.
    (g) Example of degree of coherence control. We show the reconstructed degrees of coherence ($\gamma_{\mathrm{rec}}$) with respect to the encoded ones ($\gamma_{\mathrm{rec}}$). We choose the degrees of coherence of the three pairs of fields ($\gamma_{12}$, $\gamma_{13}$ and $\gamma_{23}$) to be equal.}
    \label{fig:procedure}
\end{figure*}

\subsection{Preparation}
The transmission matrix of the complex medium must be characterized to employ it as a part of the reconfigurable multi-port linear device.
Each element of the TM connects the field modulated by the $n$th pixel of SLM1 to the complex field of the $m$th output mode (a camera pixel used for TM characterization).
We reconstruct the TM by configuring the SLM with each vector of a complete basis of the input modes, and measuring the corresponding complex fields at the output camera.
Having at our disposal phase-only SLMs, we choose the Hadamard basis, which maximizes the measured intensity and increases the signal-to-noise ratio of the reconstruction \cite{Popoff2010}.
Moreover, we measure the phase of the outputs with an interference measurement, employing part of the SLM to provide a static reference field \cite{Popoff2010, Dubois2002}.

As discussed in Sec.~\ref{sec:implemetation}A, the speckle patterns generated by each input field are mutually incoherent, thus they do not interfere.
Therefore, we assign a different transmission matrix $\hat{T}_{\mathrm{CM}, i}$ to each of the three non-overlapping  input beams.
Each laser is spatially phase modulated by a different quadrant consisting of $256\times256$ pixels of SLM1, out of a total of $512\times512$ pixels.
The outer part of each quadrant is used as a static reference for the interference measurement, while we use an area of $128\times128$ pixels (divided into $4096$ square segments of 4 pixels each) to encode the Hadamard basis employed in the TM reconstruction.
Once we have reconstructed the TM for each input laser, we can implement any desired linear transformation according to Eq.~\eqref{eq:inversion}.
We refer the reader to Supplement 1 for a thorough characterization of the multi-port linear device.

\subsection{Verification} \label{subsec:DoCrec}
The multi-port linear device described above is able to encode any desired coherence matrix.
To verify the correctness of the encoding, we measure each entry of the coherence matrix, that is the mutual degree of coherence $\gamma_{ij}$ of each field pair.
The measurement of $|\gamma_{ij}|$ can be carried out from the relations \cite{BornWolf}
\begin{subequations}\label{eq:DoCreconstruction}
\begin{eqnarray}
    |\gamma_{ij}| =  \frac{I_i + I_j}{2\sqrt{I_i I_j}} \mathcal{V} \;,\\
    \mathcal{V} = \frac{I_{\mathrm{max}} - I_{\mathrm{min}}}{I_{\mathrm{max}} + I_{\mathrm{min}}}\;.
\end{eqnarray}
\end{subequations}
where $\mathcal{V}$ is the visibility, $I_{\mathrm{max}}$ and $I_{\mathrm{min}}$ are the maximum and minimum of the interference fringes, respectively, and $I_i$ and $I_j$ are the single fields' intensities.
We highlight that all the quantities are defined at a single point in the camera plane, given that we can tune the relative phase of the interfering beams, as we discuss later.
Moreover, even if $\gamma_{ij}$ is a complex quantity, we only consider its magnitude, as a change in the phase results in a trivial shift of the interference fringes.
In the following, in writing degree of coherence $\gamma_{ij}$ we will always refer to its magnitude.

In Fig.~\ref{fig:procedure}, we summarize the procedure employed to measure the degree of coherence.
Firstly, we use SLM2 to apply a linear phase grating to two of the three output beams \cite{Davis1999}.
In the focal plane, which corresponds to the camera plane, the phase grating spatially displaces the two beams, allowing us to measure the intensity of the remaining one.
In Fig.~\ref{fig:procedure}a and Fig.~\ref{fig:procedure}b, we show the intensity distribution of the first and the second beam, respectively, when the other two are displaced.
We then use the phase grating to displace only one beam, and let the other two interfere, leading to the typical sinusoidal modulation across the area of the camera, as shown in Fig.~\ref{fig:procedure}c and Fig.~\ref{fig:procedure}d.
In particular, Fig.~\ref{fig:procedure}c shows the intensity distribution when the mutual degree of coherence of the two interfering beams is low ($\gamma_{ij} = 0.2$), while Fig.~\ref{fig:procedure}d shows the case of high degree of coherence ($\gamma_{ij} = 0.8$).
Figure \ref{fig:procedure}e shows a cross-section of the interference fringes for three degrees of coherence ($\gamma_{ij} = 0.2, 0.5, 0.8$).
The modulation depth increases for higher degrees of coherence, as expected.
Next, we modulate the phase of one of the beams (termed reference phase later) from $0$ to $2\pi$.
This modulation results in a spatial shift of the interference fringes.
Thus, we are able to measure the visibility at each pixel of the camera.
Figure \ref{fig:procedure}f shows examples of the intensity at a camera pixel with respect to the reference phase for three different degrees of coherence ($\gamma_{ij} = 0.2, 0.5, 0.8$).
From the visibility and the intensities of the single beams, we reconstruct the degree of coherence at a specific location, according to Eq.~\eqref{eq:DoCreconstruction}.
As the reconstruction of the degree of coherence is noisier for regions of low intensities, we choose to only consider pixels where both the single-beam intensities are above 60\% of their maximum value. 
We repeat the previous procedure for all the considered pixels, and average the results to obtain the reconstructed degree of coherence $\gamma_{ij}$.

Inevitably, the encoding of a chosen coherence matrix is subject to errors, leading to a discrepancy between the encoded and the reconstructed $\gamma_{ij}$.
To minimize this discrepancy, we implement a gradient descent algorithm to optimize the multi-port linear device for minimum error (see Supplement 1 for details).
The gradient descent is performed a single time to achieve better calibration of the linear port.
The phase masks obtained with this final procedure can be then employed without further calibration as long as the complex medium is stable (which in the case of ground glass diffusers is only limited by the pointing stability of the lasers \cite{Matthes:19}).

Figure \ref{fig:procedure}g shows an example of the achieved precision in the control of the degree of coherence.
Here, we report the reconstructed degrees of coherence $\gamma_{\mathrm{rec}}$, with respect to the encoded values $\gamma_{\mathrm{enc}}$.
We encoded coherence matrices with identical degrees of coherence between each field pair, i.e., $\gamma^{\mathrm{enc}}_{12}=\gamma^{\mathrm{enc}}_{13}=\gamma^{\mathrm{enc}}_{23}$, ranging from 0 to 1.
The reconstructed and the encoded degrees of coherence agree to within an average error of 0.004 in the region between 0.2 and 0.8.
Outside this range, we observe deviations from the expected behaviour. 
For low coherence, the measurement of $\gamma_{ij}$ is affected by the background noise caused by the uncontrolled modes of the complex medium, whereas for high coherence we are limited by the self-coherence of the input lasers (see Supplement 1).

\section{Results}
\begin{figure*}
    \centering
    \includegraphics[width=\textwidth]{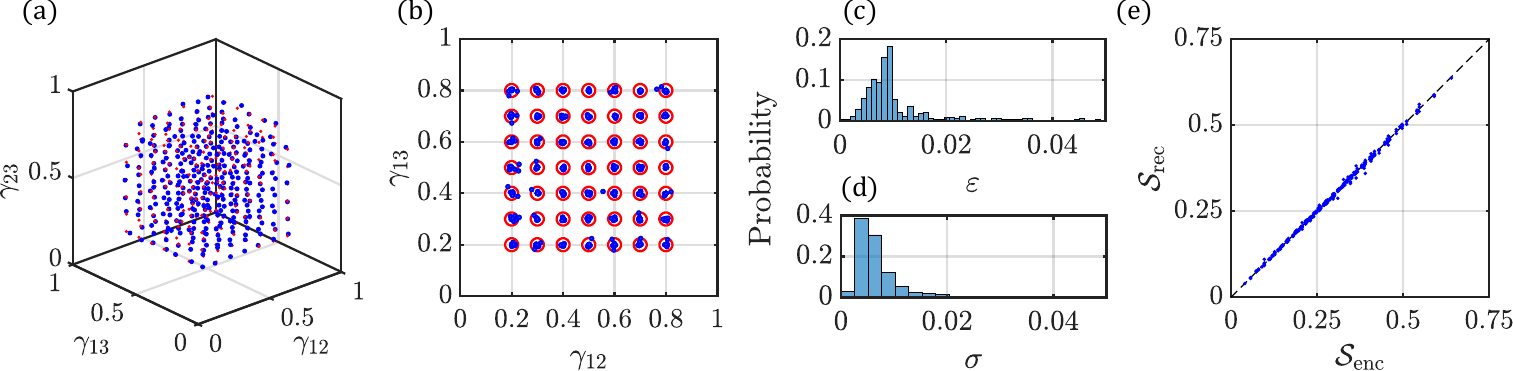}
    \caption{Coherence matrix control.
    (a) Coherence matrix space. 
    Each point of the space represents a different coherence matrix, for which the off-diagonal elements are given by the three coordinates of the point $[\gamma_{12}, \gamma_{13}, \gamma_{23}]$. 
    The blue and the red dots are the encoded and reconstructed coherence matrices, respectively.
    The encoded degrees of coherence range from 0.2 to 0.8.
    (b) One of the 2-dimensional orthographic projections of the three-dimensional space. 
    The blue dots represent the measured vectors, while the red circles with radius 0.025 are centered on the encoded vectors. 
    (c) Histogram of the error $\varepsilon$, i.e. the distance between the reconstructed and the encoded values. 
    (d) Histogram of the statistical error $\sigma$. For each encoded coherence matrix, we repeat the reconstruction of the three degrees of coherence 10 times.
    The statistical error $\sigma$ is calculated as the standard deviation of each ensemble.
    (e) Encoded overall coherence $\mathcal{S}_{\mathrm{enc}}$ vs. reconstructed one $\mathcal{S}_{\mathrm{rec}}$. The blue dots are the measured values, while the dashed line represents the ideal relation.
    }
    \label{fig:results}
\end{figure*}
We now show the level of control over the coherence matrix achievable with the presented implementation.
The coherence matrix is completely defined by its off-diagonal values, which are the mutual degrees of coherence of the field-field pairs.
We can therefore assign to each coherence matrix a vector $\pmb{\gamma} = [\gamma_{12}, \gamma_{13}, \gamma_{23}]$, and visualize all the possible vectors in a space where the axes are the magnitudes of the mutual degrees of coherence.
Note that the positive semi-definiteness of the coherence matrix bounds the domain of allowed vectors (see Supplement 1 for details).
We discretize the three-dimensional space in a cubic grid with a step size of 0.1.
In Fig.~\ref{fig:results}a, we show the experimentally achieved coherence matrices.
The blue (red) dots represent the measured (encoded) vectors.
We restrict ourselves to the region of degrees of coherence between 0.2 and 0.8, where the reconstructed coherence matrices do not have a large deviation with respect to the encoded ones, caused by technical limitations (see Fig.~\ref{fig:procedure}g and Supplement 1).
To graphically make more evident the typical distance between encoded and reconstructed coherence matrices, we show in Fig.~\ref{fig:results}b one of the orthographic projections of the three-dimensional space.
The blue dots represent the reconstructed vectors, while the red circles with a radius of 0.025 are centered on the encoded matrices, and provide a visual reference.

As a quantitative measure for the accuracy of our coherence matrix control scheme, we define, for each measured point, the error $\varepsilon$ as the root-mean-square distance between the encoded and the reconstructed vector in the space of the coherence matrices: 
\begin{equation}
    \varepsilon = \sqrt{\sum_i(\gamma_i^{\mathrm{enc}} - \gamma_i^{\mathrm{rec}})^2}\;.    
\end{equation}
Here the subscript $i$ indicates the field-field pairs $i = \{1,2\}, \{1,3\}, \{2,3\}$.
Figure \ref{fig:results}c shows the histogram of the errors.
For the majority of the coherence matrices, the error is below 0.01, which is the threshold value set in the gradient descent optimization.
The mean value of the error is 0.01.

Next, we characterize the statistical error associated with the measured $\pmb{\gamma}$.
To do so, we repeat the previously described reconstruction procedure 10 times for each vector of the space.
We then estimate the statistical error $\sigma$ as the standard deviation of the measured ensemble.
In Fig.~\ref{fig:results}d we plot the histogram of $\sigma$.
The average statistical error is 0.008. 
This justifies the chosen threshold in the optimization algorithm.

Finally, we characterize the whole system of fields with a single quantity, i.e., the overall coherence. 
The overall coherence $\mathcal{S}$ is a real number ranging from 0 (full incoherence) to 1 (full coherence) and measures the coherence of the whole system, independently of how it is shared between the degrees of coherence (see Sec.~\ref{sec:theory}A).
In Fig.~\ref{fig:results}e, we plot the reconstructed ($\mathcal{S}_{\mathrm{rec}}$) versus encoded ($\mathcal{S}_{\mathrm{enc}}$) overall coherence, computed from all the measured vectors shown in Fig.~\ref{fig:results}a.
The average error (defined as $\left| \mathcal{S}_{\mathrm{enc}} - \mathcal{S}_{\mathrm{rec}} \right|$) over all the measurements is 0.003.

\section{Discussion}
In summary, we have presented a technique to program the coherence matrix of a set of spatially separated fields through the use of a multi-port linear optical device.
We have experimentally realized a $3 \times 3$-port system, based on wavefront shaping of mutually incoherent inputs that propagate through a complex medium.
By sampling the set of allowed coherence matrices, we have shown that we can encode and successfully retrieve the majority of the matrices within an average error of 0.01.
Configurations with overall coherence close to 0 and 1 are not obtainable with our implementation.
Nevertheless, we point out that those are cases in which our scheme has no advantages compared to traditional approaches.
In fact, for low coherence one can directly employ the input fields. For high coherence, instead, one could use a single laser and split it with beam splitters.

Remarkably, to our best knowledge this is the first time that the spatial coherence of multiple fields is controlled in a single-shot fashion.
Single-shot means that, once the correct phase mask is programmed into the SLM, the spatial coherence modulation occurs after a single propagation through the system, in contrast with previous works which rely on the collection of a large ensemble of phase masks introduced by some active device \cite{DeSantis1979, Hyde2015, Rodenburg2013}.
Moreover, our complex medium-based device can operate on many ports.
In fact, in order to increase the number of controlled fields with the same performance in terms of coherence control, we only need to keep a constant background, which allows us to retain the same minimum coherence (see Supplement 1).
We can achieve this by maintaining a constant number of SLM pixels for each input laser \cite{Popoff2011b}, for instance by employing a larger SLM.

Our work adds an important tool to the available methods for controlling the various attributes of light.
Among the various potential applications \cite{Korotkova2020}, our work is of considerable interest for free-space optical communications, where beams with partial spatial coherence are more robust to atmospheric turbulence compared to the coherent counterpart \cite{Korotkova2004, Gbur2014}.
Moreover, one could use the mutual degree of coherence as a physical bit, leading to a favorable quadratic scaling of the number of communication channels with the number of input fields. 

\paragraph{Funding.} ETH Zurich Research Grant (ETH-41 19-1); Jane and Aatos Erkko Foundation; H2020 European Research Council (SMARTIES-724473).

\paragraph{Disclosures.} The authors declare that there are no conflicts of interest related to this article.

\paragraph{Data availability.} The data that support the findings of this study are available from the corresponding author upon reasonable request.

\paragraph{Supplemental document.} See Supplement 1 for supporting content.

\bibliography{references}

\include{supplemental/supplemental}

\end{document}

%% file: supplemental/supplemental.tex
\setcounter{figure}{0}
\setcounter{equation}{0}
\setcounter{section}{0}
\renewcommand{\thefigure}{S\arabic{figure}}
\renewcommand{\theequation}{S\arabic{equation}}

\begin{center}
	\textbf{\LARGE Controlling spatial coherence with an optical complex medium: supplemental document} \\ 
\end{center}
\vspace{3mm}

\section{Technical limitations}\label{app:limitations}
In the main text, we show that we encounter deviations between the encoded and the reconstructed degree of coherence when we try to achieve very low (approaching 0) or very high (approaching 1) values.
In this section, we provide more details on the origins of these limits.

Let us start investigating the case where we want to obtain mutually incoherent outputs.
According to Eq.~(7) in the main text, the linear transformation we want to apply is the identity matrix, i.e., the inputs should be transmitted to the outputs unaffected.
However, since it is not feasible to control all the modes supported by the scattering medium, the beams get mixed during the propagation through the medium.
The resulting background noise is responsible for an unwanted contribution of each input field to every output.

\begin{figure}
    \centering
    \includegraphics[width=0.75\textwidth]{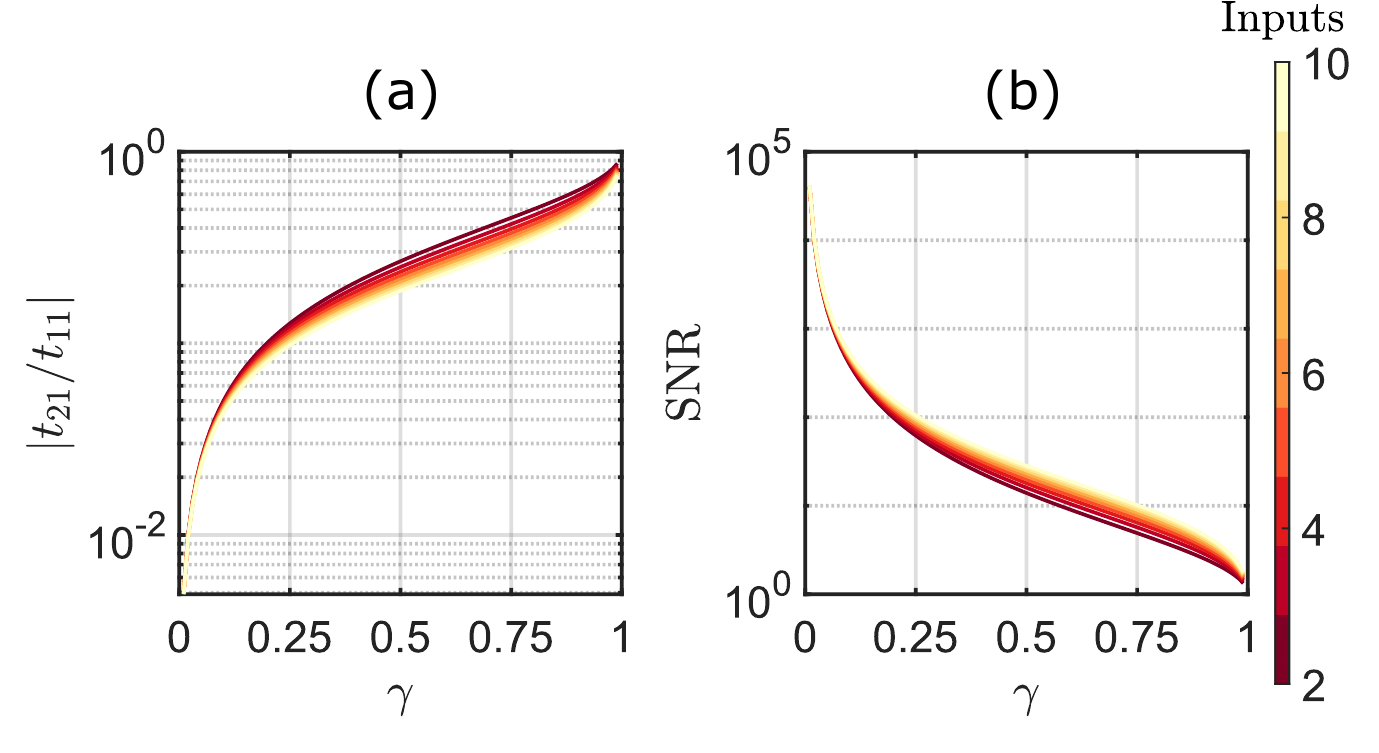}
    \caption{Minimum degree of coherence limitations.
    (a) Given two outputs $E_1^{\mathrm{out}} = t_{11} E_1^{\mathrm{in}}$ and $E_2^{\mathrm{out}} = t_{21} E_1^{\mathrm{in}}$, the degree of coherence $\gamma$ depends on the ratio $|t_{21}/t_{11}|$. 
    (b) Minimum signal-to-noise ratio (SNR) needed to encode the degree of coherence $\gamma$. In the case that $|t_{21}|$ is only given by the background noise, the SNR is $|t_{11}|^2/|t_{21}|^2$ .}
    \label{fig:minimum_coherence}
\end{figure}

We will now quantify the limitations to the minimum degree of coherence imposed by the background noise.
We start considering a coherence matrix $\mathbb{K}_{\mathrm{out}}$ of the form
\begin{equation}\label{eq:chosenK}
    \mathbb{K}_{\mathrm{out}} = 
    \begin{bmatrix}
        1 & \gamma & \cdots & \gamma \\
        \gamma & 1 & \cdots & \gamma \\
        \vdots & \ddots &\ddots & \vdots \\
        \gamma & \gamma & \cdots & 1
    \end{bmatrix}\;,
\end{equation}
where, for simplicity, we set the off-diagonal terms to have the same constant real value $\gamma$.
To have mutually incoherent output fields, we want $\gamma$ to tend to $0$.
From $\mathbb{K}_{\mathrm{out}}$ we extract the expression of the linear transformation $\hat{T}$ [Eq.~(7) in the main text], which connects the mutually incoherent input fields $\boldsymbol{E}_{\mathrm{in}}$ to the output fields $\boldsymbol{E}_{\mathrm{out}}$ with coherence matrix $\mathbb{K}_{\mathrm{out}}$.
%
%
From the form we chose for $\mathbb{K}_{\mathrm{out}}$ [Eq.~(\ref{eq:chosenK})], the linear transformation $\hat{T}$ can be completely described by two coefficients: $t_{11}$ for the diagonal terms, which are all equal, and $t_{21}$ for the off-diagonal elements, which are again all equal.
Note that the coefficients $t_{11}$ and $t_{21}$ associate the two outputs $E_1^{\mathrm{out}}$ and $E_2^{\mathrm{out}}$ with the single input $E_1^{\mathrm{in}}$, according to the relations $E_1^{\mathrm{out}} = t_{11}E_1^{\mathrm{in}}$, and $E_2^{\mathrm{out}} = t_{21}E_1^{\mathrm{in}}$ [Eq.~(8) in the main text].
Ideally, we would like $|t_{21}|$ to approach zero to get zero output degree of coherence, i.e., we want $|E_2^{\mathrm{out}}| = 0$.
In practice, the background noise in the output intensity pattern poses a lower bound to the intensity $|E_2^{\mathrm{out}}|^2$, hence to $|t_{21}|$, which finally sets the minimum degree of coherence different from zero.
In Fig.~\ref{fig:minimum_coherence}a, we show the scaling of the absolute value of the ratio $|t_{21}/t_{11}|$ as a function of the degree of coherence $\gamma$.
If we increase the number of inputs, i.e., the dimensionality of $\mathbb{K}_{\mathrm{out}}$, the requirement is very similar (Fig.~\ref{fig:minimum_coherence}).
A desired minimum degree of coherence translates into a minimum signal-to-noise ratio (SNR). 
In fact, considering the single input $E_1^{\mathrm{in}}$ and assuming that the only contribution to $E_2^{\mathrm{out}}$ is given by the background noise, $|t_{11}|^2$ is the maximum generated intensity and $|t_{21}|^2$ is the noise intensity, thus the SNR is defined as $|t_{11}|^2/|t_{21}|^2$.
In Fig.~\ref{fig:minimum_coherence}b, we show that low coherence values demand very high SNR, which is limited by the number of SLM pixels modulating each input laser \cite{supp:Popoff2011b}.
A similar argument works for a different form of $\mathbb{K}_{\mathrm{out}}$, where the limitation is given by the element of the transformation $\hat{T}$ with the minimum absolute value.

\begin{figure}
    \centering
    \includegraphics[width=\textwidth]{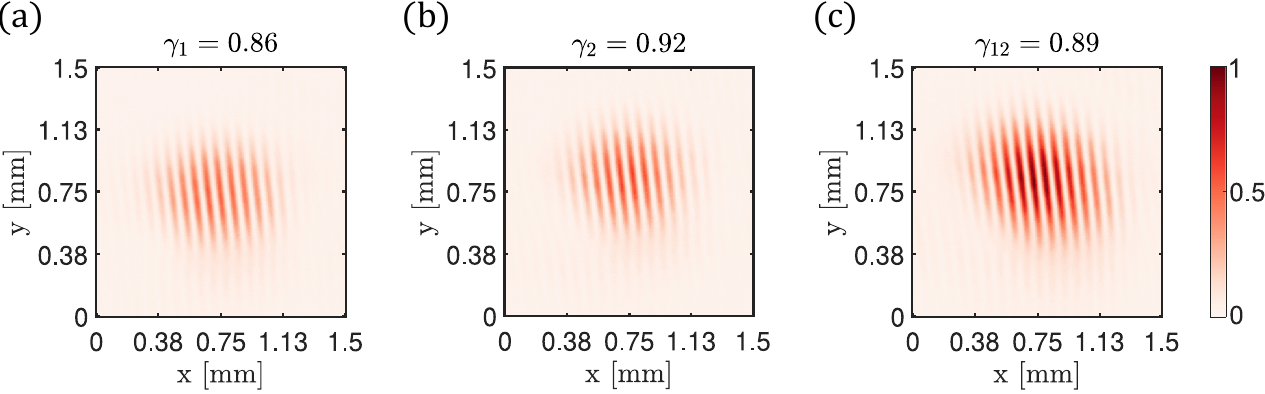}
    \caption{Interference patterns. We have two interfering output fields, resulting from the superposition of two inputs. We show the interference patterns when (a) the first, (b) the second or (c) both inputs contribute to the outputs. Each case is associated to a measured degree of coherence $\gamma_i$, where the subscript $i$ indicates the contributing inputs.}
    \label{fig:maximum_visibility}
\end{figure}

Let us now consider the factors limiting the maximum degree of coherence.
To investigate this case, we turn on a single input ($E_1^{\mathrm{in}}$), and we consider a single pair of output fields $E_1^{\mathrm{out}}$ and $E_2^{\mathrm{out}}$, which are related to the input by the coefficients $t_{11}$ and $t_{21}$, as discussed above.
We compute the mutual degree of coherence
\begin{equation}
\gamma = \frac{\langle E_1^{\mathrm{out}}\left(E_2^{\mathrm{out}}\right)^{\ast}\rangle}{\sqrt{\langle|E_1^{\mathrm{out}}|^{2}\rangle\langle|E_2^{\mathrm{out}}|^{2}\rangle}}
=  \frac{t_{11} t_{21}^* \left\langle |E_1^{\mathrm{in}}|^2 \right\rangle}{|t_{11} t_{21}| \left\langle |E_1^{\mathrm{in}}|^2 \right\rangle}  = \frac{t_{11} t_{21}^*}{|t_{11} t_{21}|}\;,
\end{equation}
whose modulus $|\gamma|$ is always equal to 1, regardless the values of the transformation coefficients.
Nevertheless, the measurements deviate from this ideal result.
To show it, we use a single input laser to generate through our system two output beams.
We then let the output beams interfere and we reconstruct the degree of coherence.
We report the measured interference patterns for two different input lasers in Fig.~\ref{fig:maximum_visibility}a and \ref{fig:maximum_visibility}b.
The reconstructed degrees of coherence ($\gamma_{1} = 0.86$ for the first input and $\gamma_{1} = 0.92$ for the second) are lower than the ideal value of $1$.
This discrepancy, in line with what is reported in literature, is associated to the limited spatial coherence of the light source \cite{supp:Leppanen2017}.
We show now that the maximum degree of coherence achievable with a single laser is limiting the value obtainable by the whole system.
Let us consider two mutually incoherent inputs, both of them contributing to two output fields.
Since the components from the different inputs do not interfere, the resulting interference pattern is given by the sum of the individual patterns.
Thus, we can write the visibility in terms of the maximum $I^{\mathrm{max}}_1$, $I^{\mathrm{max}}_2$ and minimum $I^{\mathrm{min}}_1$, $I^{\mathrm{min}}_2$ intensity given by the contributions from the two different inputs:
\begin{equation}
\mathcal{V} = \frac{(I^{\mathrm{max}}_1 + I^{\mathrm{max}}_2) - (I^{\mathrm{min}}_1 + I^{\mathrm{min}}_2)}
{(I^{\mathrm{max}}_1 + I^{\mathrm{max}}_2) + (I^{\mathrm{min}}_1 + I^{\mathrm{min}}_2)}\;.
\end{equation}
After few algebraic passages, we get
\begin{equation}
\mathcal{V} = \frac{\mathcal{V}_1}{1 + \frac{(I^{\mathrm{max}}_2 + I^{\mathrm{min}}_2)}{(I^{\mathrm{max}}_1 + I^{\mathrm{min}}_1)}} + \frac{\mathcal{V}_2}{1 + \frac{(I^{\mathrm{max}}_1 + I^{\mathrm{min}}_1)}{(I^{\mathrm{max}}_2 + I^{\mathrm{min}}_2)}}\;,
\end{equation}
where $\mathcal{V}_i = (I^{\mathrm{max}}_i - I^{\mathrm{min}}_i)/(I^{\mathrm{max}}_i + I^{\mathrm{min}}_i)$ is the visibility of the interference pattern given by the $i$th input.
Considering $I_1^{\mathrm{max}} = I_2^{\mathrm{max}}$ and $I^{\mathrm{max}}_{1,2} \gg I^{\mathrm{min}}_{1,2}$, we obtain
\begin{equation}\label{eq:max_visibility}
\mathcal{V} \approx \frac{\mathcal{V}_1 + \mathcal{V}_2}{2}\;.
\end{equation}
The last equation tells us that the maximum visibility obtainable by the whole system [directly linked to the degree of coherence, see Eq.~(11a) in the main text] is limited by the average visibility over each single input.
Therefore, the maximum degree of coherence achievable is limited by the spatial coherence of the light sources.
Figure \ref{fig:maximum_visibility} shows the interference pattern when we turn on: (a) only the first input, (b) only the second one, or (c) both of them.
The measured degrees of coherence resulting from the combination of the two inputs ($\gamma_{12} = 0.89$) is in agreement with Eq.~(\ref{eq:max_visibility}).

\section{Mutual incoherence of the input fields}
\begin{figure}
    \centering
    \includegraphics[width=\textwidth]{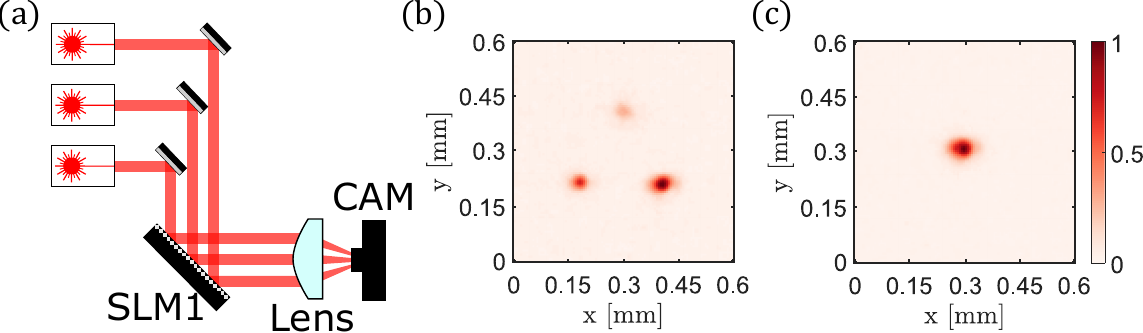}
    \caption{Mutually incoherent inputs. (a) Characterization setup. Three independent laser beams are modulated by a SLM before being focused onto a camera. (b, c) Camera images. The SLM is used to (b) separate the beams in the focal plane or to (c) focus them in the same point. No interference fringes are present when the beams overlap, confirming that the three fields are mutually incoherent.}
    \label{fig:incoherent_inputs}
\end{figure}

Our implementation relies on mutually incoherent inputs.
To achieve this condition, we used three red lasers (Thorlabs HRP050 and Meredith Instruments $633\,\mathrm{nm}$ HeNe lasers, and $\approx 650\,\mathrm{nm}$ pen-type visual fault locator FOSCO BOB-VFL650-10), with a linewidth (HeNe $\approx 10\,\mathrm{MHz}$, VFL $\approx 1\,\mathrm{THz}$) much larger than the bandwidth of the employed detector (Basler acA640-750um, bandwidth $\approx 10 \div 100\,\mathrm{Hz}$). 
This ensures that we can consider them mutually incoherent.
To confirm it, we focused the three laser beams into a single spot, checking that no interference fringes are visible (see Fig.~\ref{fig:incoherent_inputs}).

\section{Coherence matrix space}
\begin{figure}
    \centering
    \includegraphics[width=0.75\textwidth]{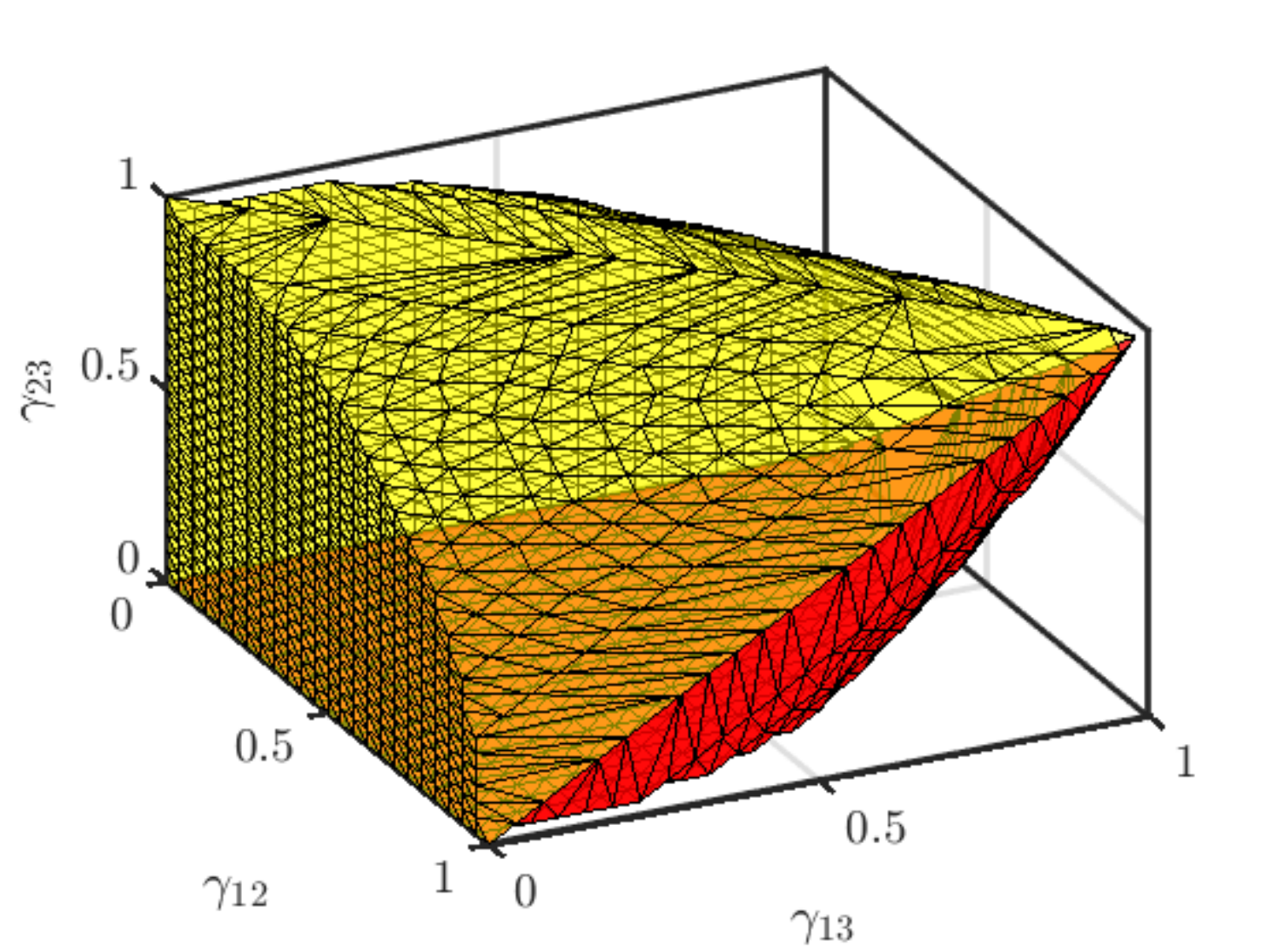}
    \caption{Graphical representation of the space of allowed coherence matrices for three fields $E_1$, $E_2$ and $E_3$. $\gamma_{ij}$ represents the mutual degree of coherence of the two fields $E_i$ and $E_j$.}
    \label{fig:3Dspace_coherence}
\end{figure}

In this section we describe the space of the allowed coherence matrices.
As discussed in the main text, the coherence matrix is Hermitian and normalized such that its diagonal elements are 1. 
For three field $E_1$, $E_2$ and $E_3$, these conditions leave three degrees of freedom, which are the mutual degrees of coherence $\gamma_{12}$, $\gamma_{13}$ and $\gamma_{23}$.
Even though the degrees of coherence are complex quantities, in this work we focus specifically on their magnitudes, as the phase corresponds only to a spatial shift of the interference fringes. 
Thus, for three fields, we can visualize all possible combinations of the degree of coherence magnitudes in a three-dimensional space.
However, not any combination of degrees of coherence is physically acceptable.
In fact, the coherence matrix must also be positive semi-definite \cite{supp:Devroye1986}.
It translates into the condition of real and positive (or zero) eigenvalues.
We show in Fig.~\ref{fig:3Dspace_coherence} the contour of the space where the positive semi-definiteness is satisfied.

\section{Linear port}
In this section we characterize the linear transformation implemented with the system of complex medium and SLM.
In the main text, we focused on the specific case of a $3\times 3-$port, that is a device that mixes three input fields into three output fields with controlled amplitude and phase.
Thus, each input $E^{\mathrm{in}}_i$ generates three outputs according to $E^{\mathrm{out}}_1 = t_{1i} E^{\mathrm{in}}_i$, $E^{\mathrm{out}}_2 = t_{2i} E^{\mathrm{in}}_i$ and $E^{\mathrm{out}}_3 = t_{3i} E^{\mathrm{in}}_i$.
Turning off two of the three inputs, we can experimentally measure the coefficients $t_{1i}$, $t_{2i}$ and $t_{3i}$.
Characterizing these coefficients is the topic of the present section.

\begin{figure}
    \centering
    \includegraphics[width=0.75\textwidth]{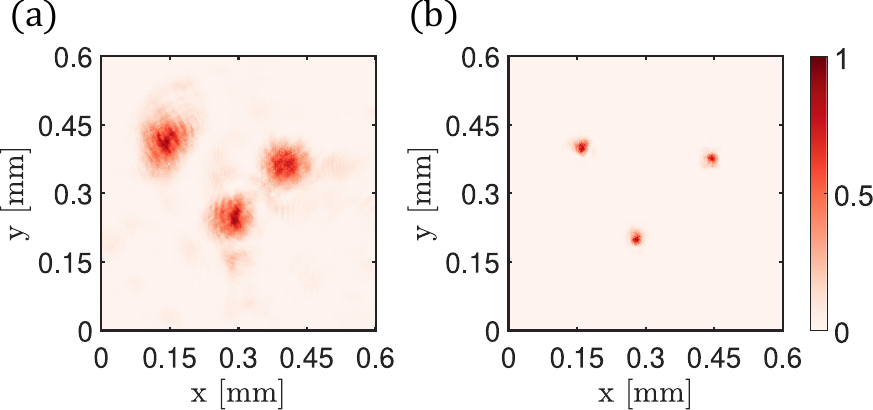}
    \caption{Spatial filtering. Output fields' intensity distributions (a) before and (b) after three small circular apertures placed before the second SLM.}
    \label{fig:pinholes_figure}
\end{figure}

We consider the input $E^{\mathrm{in}}_1$, and we measure the output intensities $I_1 = |t_{11} E^{\mathrm{in}}_1|^2$, $I_2 = |t_{21} E^{\mathrm{in}}_1|^2$ and $I_3 = |t_{31} E^{\mathrm{in}}_1|^2$.
We show an example of the resulting intensity distributions in Fig.~\ref{fig:pinholes_figure}a.
The output beams, resulting from a speckle pattern, do not show a clean Gaussian profile.
This is detrimental for the reconstruction of the degree of coherence from the interference pattern.
Therefore, we introduce three small circular apertures ($0.5\,\mathrm{mm}$ in diameter, spaced by roughly $2\,\mathrm{mm}$) before the second SLM.
We show in Fig.~\ref{fig:pinholes_figure}b the resulting spatially filtered beams.

\begin{figure}
    \centering
    \includegraphics[width=\textwidth]{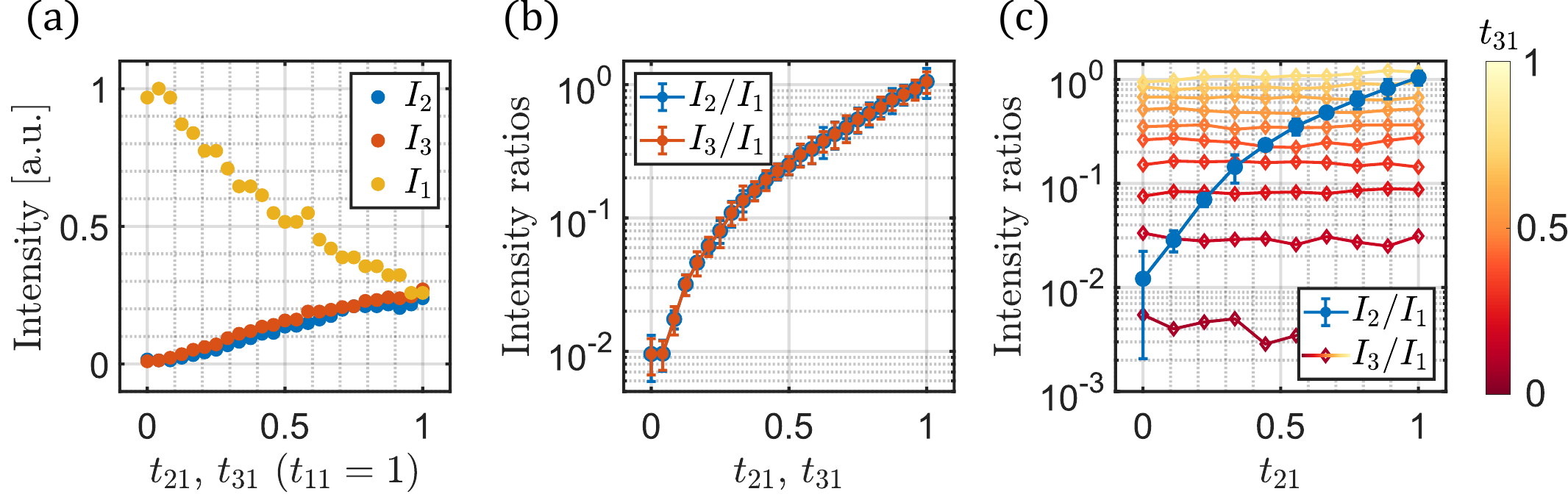}
    \caption{Linear port characterization.
    (a) Power conservation. If we increase $t_{21}$ and $t_{31}$, keeping constant $t_{11}$, the amplitude of the high intensity output reduces, to conserve the overall power shared between the outputs. 
    (b) Intensity ratios. We measured the ratios $I_2/I_1$ and $I_3/I_1$ for increasing $t_{21}$ and $t_{31}$.
    The error bars (which show the maximum deviation from the mean value) are obtained repeating the measurement 100 times.
    (c) Cross-talk analysis. We modulate $t_{21}$ from $0$ to $1$, while keeping $t_{31}$ constant. We repeat the measurement changing the value of $t_{31}$. The error bars on $I_2/I_1$ show the maximum deviation from the mean value.}
    \label{fig:TF_figure}
\end{figure}

We then characterize the output intensities when we modify the encoded coefficients.
Given the desired coefficients, we calculate the needed SLM mask $\hat{T}_{\mathrm{SLM}, i}$ according to Eq.~(10) in the main text.
We then increase the magnitudes of $t_{21}$ and $t_{31}$ from 0 to 1, keeping $t_{11}$ constant and equal to 1 (Fig.~\ref{fig:TF_figure}a).
We measure that $I_1$ decreases while we increase the intensities $I_2$ and $I_3$.
This happens mainly because the overall power distributed in the three outputs is conserved between the transformations.
Thus, if we increase the intensities of the second and third output, then $I_1$ must decrease accordingly.
We correct for this effect by characterizing the intensity ratios $I_2/I_1$ and $I_3/I_1$, which are the relevant quantities for the linear port (Fig.~\ref{fig:TF_figure}b), and we use the measured characteristics to calibrate the SLM masks.
We repeat the measurement 100 times (for a total time of about 30 minutes), resulting in the reported error bars, which show the maximum deviation from the mean value.

The next step is to characterize the cross-talk between the output beams. 
In fact, if the outputs are not completely independent, changing the intensity of one of them will affect the other two.
In Fig.~\ref{fig:TF_figure}c we increase the intensity of the output $I_2$ ($t_{21}$ from 0 to 1), while keeping $I_3$ constant.
We then repeat the measurement increasing the magnitude of $t_{31}$.
We find that the fluctuations of the intensity $I_3$ are within the error bar of $I_2$, which is comparable to the typical error that we report in Fig.~\ref{fig:TF_figure}b.
We then conclude that the systematic cross-talk is below the statistical noise, hence not relevant.

\section{Gradient descent optimization}
\begin{figure}
    \centering
    \includegraphics[width=0.75\textwidth]{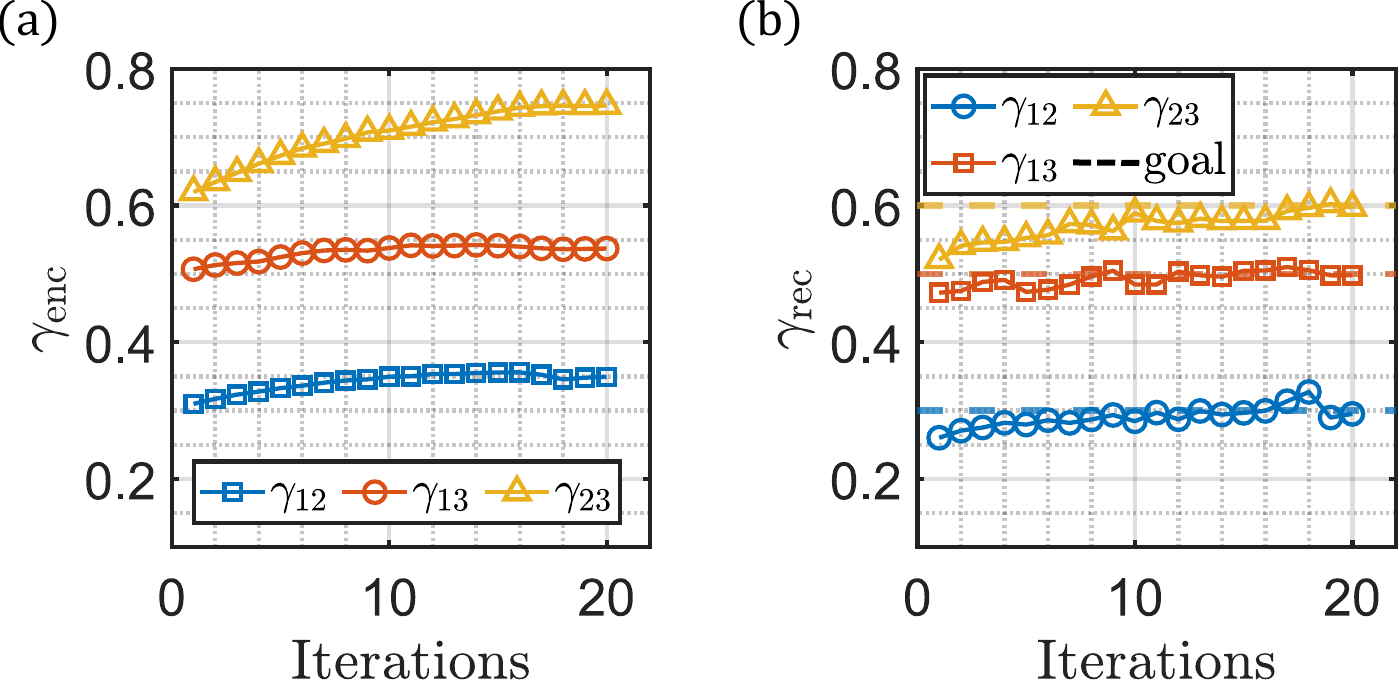}
    \caption{Feedback. (a) The encoded values $\gamma_{\mathrm{enc}}$ are iteratively corrected using the error between the desired and the measured degrees of coherence. (b) Consequently, the reconstructed degrees of coherence $\gamma_{\mathrm{rec}}$ converge to the desired values, which in this case are $\gamma_{12} = 0.3$, $\gamma_{13} = 0.5$ and $\gamma_{23} = 0.6$.}
    \label{fig:feedback}
\end{figure}
In this section we describe the feedback mechanism employed to achieve the accuracy in the control of the coherence matrix reported in the main text.
Mainly due to errors in the calibration of the SLM phase masks [$\hat{T}_{\mathrm{SLM}, i}$ in Eq.~(10) of the main text], we measure deviations between the reconstructed and the encoded degrees of coherence.
Thus, we use a gradient descent algorithm to minimize the encoding errors.

Let us consider the pair of fields $E_i$ and $E_j$.
At the $n$th iteration step, we encode the degree of coherence $\gamma_{ij}^{\mathrm{enc}}(n)$, and reconstruct $\gamma_{ij}^{\mathrm{rec}}(n)$.
We evaluate the encoding error $\varepsilon(n) = \gamma_{ij}^{\mathrm{rec}}(n) - \gamma_{ij}^{\mathrm{enc}}(n)$ and, for the next iteration, we  correct the encoded value following the relation
\begin{equation}
    \gamma_{ij}^{\mathrm{enc}}(n+1) =\gamma_{ij}^{\mathrm{enc}}(n) + \eta \varepsilon(n) \;,
\end{equation}
where $\eta$ is the feedback strength, that we used for all the coherence matrices.
From the new values of $\gamma_{ij}^{\mathrm{enc}}(n+1)$ for each pair, we construct the corrected linear port.
We reiterate the process until we are satisfied with the final encoding error ($\varepsilon < 0.01$ in our case).
The gradient descent is the last step of the calibration of the setup.
After running it once, we know the coefficients of the linear port which minimize the error, and the reconstructed degree of coherence is stable over time.

We illustrate the optimization procedure in Fig.~\ref{fig:feedback}.
We want to encode the values $\gamma_{12} = 0.3$, $\gamma_{13} = 0.5$ and $\gamma_{23} = 0.6$.
At each iteration, we correct the encoded values (Fig.~\ref{fig:feedback}a), while the reconstructed degree of coherence converge to the desired quantity (Fig.~\ref{fig:feedback}b).